\documentclass[aps,eqsecnum,showpacs,preprint,nofootinbib]{revtex4}
\usepackage{color,graphicx}
\usepackage{amsfonts}
\usepackage{amssymb}
\usepackage{amsmath}

\begin{document}

{\baselineskip0pt
\leftline{\large\baselineskip12pt\sl\vbox to0pt{\hbox{\it Department of
Mathematics and Physics}\hbox{\it Osaka City  University}
\hbox{Yukawa Institute for Theoretical Physics}\vss}}
\rightline{\large\baselineskip12pt\rm\vbox to20pt{\hbox{OCU-PHYS-339}
            \hbox{YITP-10-79} \hbox{AP-GR-83}
\vss}}%
}

\title{
Perturbative Analysis of a Stationary Magnetosphere in an Extreme Black Hole Spacetime\\
{\it -- On the Meissner-like Effect of an Extreme Black Hole --}
}

\author{
$^{1}$Yohsuke Takamori\footnote{Electronic address: takamori@sci.osaka-cu.ac.jp},
$^{1}$Ken-ichi Nakao\footnote{Electronic address: knakao@sci.osaka-cu.ac.jp}, \\
$^{1}$Hideki Ishihara\footnote{Electronic address: ishihara@sci.osaka-cu.ac.jp},
$^{1}$Masashi Kimura\footnote{Electronic address: mkimura@sci.osaka-cu.ac.jp}
and 
$^{2}$Chul-Moon Yoo\footnote{Electronic address: yoo@yukawa.kyoto-u.ac.jp}
}
\affiliation{
$^{1}$Department of Mathematics and Physics, Graduate School of Science, 
Osaka City University, Osaka 558-8585, Japan\\
$^{2}$Yukawa Institute for Theoretical Physics, Kyoto University, Kyoto 606-8502, Japan
}
\date{\today}

\begin{abstract}
It is known that the Meissner-like effect is seen in a
magnetosphere without an electric current in black hole spacetime: 
no non-monopole component of magnetic flux penetrates 
the event horizon if the black hole is extreme. In this paper, 
in order to see how an electric current affects the Meissner-like effect, 
we study a force-free electromagnetic system in a static 
and spherically symmetric extreme black hole spacetime. 
By assuming that the rotational angular velocity of the magnetic 
field is very small, 
we construct a perturbative solution for the Grad-Shafranov equation, which 
is the basic equation to determine a stationary, axisymmetric 
electromagnetic field with a force-free electric current. 
Our perturbation analysis reveals that, if an electric current 
exists, higher multipole 
components may be superposed upon the monopole component 
on the event horizon, even if the black hole is extreme.

\end{abstract}

\pacs{04.20.Cv, 04.20.Jb, 04.70.Bw}

\maketitle


\vskip1cm
\section{Introduction}
It is widely believed that there are supermassive black holes at the centers of
galaxies, and these are hypothesized to be the central engines 
for active galactic nuclei (AGNs) and gamma ray bursts (GRBs).
Two main possibilities are considered as the energy source.
One is the gravitational energy of accreting matter and 
the other is the rotational energy of the black hole or 
the accretion disk surrounding it.
However, the details of 
the energy extraction process are not clear. 
It is also not understood well how the energy is converted into that of 
AGNs or GRBs.

Blandford and Znajek showed that the rotational energy of a rotating black hole
can be extracted in the form of Poynting flux along magnetic field lines 
penetrating the event horizon~\cite{Blandford:1977}, which is known as the Blandford-Znajek (BZ) mechanism.
Its efficiency depends on the 
rotational velocity of the black hole and the configuration of the 
magnetic field: 
the extraction of the rotational energy becomes more efficient 
the more magnetic field lines penetrates the event horizon and the more rapidly 
the black hole rotates. 
In the BZ mechanism, poloidal magnetic fields which 
penetrate the event horizon play a crucial role  
for the energy extraction as well as for the formation of jets associated with AGNs.
In fact, some numerical studies reported that Poynting-dominated 
jets were produced~\cite{McKinney:2004,McKinney:2005,Hawley:2006}.

Bi\u{c}\'ak and Jani\u{s} showed that a 
magnetic field without an electric current is expelled from the event horizon
of a maximally rotating black hole~\cite{Bicak:1985}.
This is analogous to the Meissner effect in a superconductor. 
This effect for a rapidly rotating black hole would decrease
the efficiency of the BZ mechanism, though 
the larger rotational velocity of the black hole
would increase the efficiency.
In realistic astrophysical cases, however, 
there would be plasma around the black hole. 
How the Meissner-like effect is affected by the existence of plasma is the main subject of this paper. 
We clarify the effect of an electric current 
on the Meissner-like effect of an extreme 
black hole. 
Komissarov and McKinney studied numerically the Meissner-like 
effect of a Kerr black hole~\cite{Komissarov:2007}. 
They carried out numerical simulations for a highly conductive magnetosphere  
until it  almost reaches steady state, 
and there was no sign of the Meissner-like effect in their numerical results. 
In this paper, we study how an electric current 
affects the Meissner-like effect by solving a stationary problem analytically.

Since realistic situations are, in general, very complicated, it is 
difficult to model them. In order to reveal the essence of the plasma 
effect, we consider a very simple toy model: 
(i) we consider a stationary, axisymmetric force-free 
system of the electromagnetic field and plasma; 
(ii) we consider a static spherically 
symmetric black hole spacetime with 
a degenerate horizon as a background spacetime rather than a 
rotating black hole. 
The degenerate horizon is the origin of the Meissner-like effect in a vacuum 
black hole spacetime~\cite{Bicak:2006}, and hence, 
by studying the electromagnetic field in this spacetime, 
we can see whether the Meissner-like effect remains even in the case 
with an electric current.
The spacetime considered in this paper is known as the Reissner-Nordstr\"om (RN) spacetime. 
By these assumptions, the basic equations reduce to only one  
quasi-linear elliptic equation for the magnetic flux function
called the Grad-Shafranov (GS) equation~\cite{Macdonald:1982}. 

For the black hole spacetime, the GS equation has three regular singular points: 
one is at the event horizon, and the other two are at 
the inner and outer light surfaces 
on which the velocities of the magnetic 
field lines agree with the speed of light. 
For non-extreme cases, one boundary condition is imposed at each 
regular singular point so that the magnetic field is smooth everywhere. 
However, for a given electric current function, 
the obtained solution for the magnetic flux need not be 
$C^1$ but at most $C^{1-}$~\cite{Contopoulos:1999}.
Although numerical $C^1$ solutions have been 
obtained by iteratively changing the functional form of the 
electric current~\cite{Contopoulos:1999,Ogura:2003,Uzdensky:2004,Uzdensky:2005,Gruzinov:2005,Timokhin:2006}, 
a mathematically rigorous proof for the 
existence of a $C^1$ solution has not yet been presented. 
Furthermore, in the extreme case, 
two kinds of boundary condition must be imposed at once on the event horizon. 
We shall mention all these difficulties in solving the GS equation in \S IV. 

As will be shown in \S V,
the monopole component is a unique configuration of the magnetic field 
on the event horizon if there is not an electric current. 
Since there is no magnetic monopole in nature, 
this result implies the Meissner-like effect of the extreme RN black hole. 
In order to study the electromagnetic field 
coupled to an electric current around an RN black hole, 
we use a perturbative method which includes two expansion parameters.
One of these parameters corresponds to the rotational angular velocity of the 
magnetic fields. 
Namely, we consider slow-rotating magnetic fields as was first
considered by Blandford and Znajek~\cite{Blandford:1977}. 
The other parameter is the ratio of the distance from 
the event horizon to the horizon radius, 
since we consider only the vicinity of the event horizon, 
which includes the inner light surface. 
Although we cannot take into account the outer light surface in 
our perturbative method, 
we can obtain approximate solutions sufficient  
to study the Meissner-like effect with an electric current. 

This paper is organized as follows. 
In \S II, we introduce the RN black hole as a background geometry. 
Then we show the GS equation for the RN spacetime in \S III; the detailed derivation of 
the GS equation is given in Appendices A and B.  
The regularity conditions for the GS equation 
and difficulties in solving this equation are 
described in detail in \S IV. 
Using perturbative analyses, we study the cases with and without an electric current 
in \S V and VI, respectively. 
\S VII is devoted to summary and discussion. In Appendix C, 
we show the relation between the Kerr-Schild coordinate system and the 
standard static coordinate system of the RN spacetime. 
In Appendix D, we give a proof of a theorem on the magnetic 
field obtained by the present perturbative method. 

In this paper, we adopt the geometrized units, in which the Newton's gravitational 
constant and the speed of light are unity, and the 
abstract index notation: small Latin indices, excluding $t$ and $r$, 
indicate the type of tensor, whereas 
small Greek indices, excluding $\theta$ and $\varphi$, represent 
components with respect to 
the coordinate basis. The exceptional indices $t$, $r$, $\theta$, and $\varphi$  
denote the components of time, and the radial and azimuthal coordinates  
in the spherical polar coordinate system. 
The signature of the metric is diag$[-,+,+,+]$.  
\section{Background geometry}

We consider a static and spherically symmetric spacetime of the following metric:  
\begin{eqnarray}
ds^2=g_{\mu\nu}dx^\mu dx^\nu=-\alpha^2dt^2+\frac{r^2}{\Delta}dr^2+
r^2\left(d\theta^2+\sin^2\theta d\varphi^2\right) 
\label{metric}
\end{eqnarray}
with 
\begin{equation}
 \Delta=(r-r_+)(r-r_-)~~~~~{\rm and}~~~~~\alpha=\frac{\sqrt{\Delta}}{r},
\end{equation}
where we assume $r_+\geq r_->0$. This spacetime is known as the RN 
spacetime. There are two horizons, which are determined by $\Delta=0$: 
$r_+$ and $r_-$ represent the radius of the event and Cauchy horizons, respectively.
The case of $r_{+}=r_{-}:=r_{\rm H}$ is called the extreme case. 

\section{Grad-Shafranov equation}
Maxwell's equations are given by
\begin{eqnarray}
 \nabla_{[a}F_{bc]} &=& 0, \label{eq:Maxwell-eq-1}\\
 \nabla_bF^{ab} &=& 4\pi J^{a}, \label{eq:Maxwell-eq}
\end{eqnarray}
where $F_{ab}$ is the field strength tensor of the electromagnetic field, 
$J^a$ is a current density, and $\nabla_b$ is 
the covariant derivative\footnote{We only use the RN spacetime as a 
background geometry rather than the Kerr spacetime, that is, a test U(1) 
field on the RN spacetime is considered in our toy model.
Bi\u{c}\'{a}k and Dvo\u{r}\'{a}k studied
perturbations of the coupled Einstein-Maxwell system on the extreme RN spacetime
without an electric current~\cite{Bicak:1980}. The Meissner-like effect appeared in this case.}.
If the field strength tensor is expressed by using a 4-vector potential $A_a$ as
\begin{equation}
F_{ab}=\nabla_aA_b-\nabla_b A_a
=\partial_a A_b-\partial_b A_a, \label{eq:F-def} 
\end{equation}
then Eq.~(\ref{eq:Maxwell-eq-1}) is trivially satisfied,  
where $\partial_a$ is the ordinary derivative. 

As mentioned in $\S$I, hereafter, we consider the axisymmetric 
and stationary electromagnetic field in the RN spacetime. 
In order to make the problem simple, we assume that the system 
field satisfies the force-free condition
\begin{equation}
F_{ab}J^b=0. \label{eq:ff-condition}
\end{equation}
Hereafter, we focus on the system of only Eqs.~(\ref{eq:Maxwell-eq}) 
and (\ref{eq:ff-condition}). 
The formulation of a force-free electrodynamics field in the black hole spacetime 
was given by Macdonald and Thorne~\cite{Macdonald:1982}. 
Our formulation is based on their work.

In the case of a  stationary, axisymmetric electromagnetic 
field, we can define the ``angular velocity'' of the 
magnetic field as
\begin{equation}
\Omega_{\rm F}:=\frac{F_{t\theta}}{F_{\theta\varphi}}
=\frac{F_{tr}}{F_{r\varphi}}. 
\label{eq:OmegaF-def}
\end{equation}
The reason why $\Omega_{\rm F}$ can be regarded as the angular velocity of the 
magnetic field is described in Appendix \ref{Derivation}.
{}From Eqs.~(\ref{eq:Maxwell-eq}) and (\ref{eq:ff-condition}), the GS equation is obtained as
\begin{equation}
\Psi''+\frac{1}{\Delta}L_\theta \Psi + \frac{U}{D}+\frac{W}{\Delta D}=0,
\label{eq:GS-eq}
\end{equation}
where a prime $'$ represents a derivative with respect to $r$, 
\begin{equation}
L_\theta\Psi=\sin\theta\frac{\partial}{\partial\theta}
\left(
\frac{1}{\sin\theta}\frac{\partial\Psi}{\partial\theta}
\right),
\end{equation}
\begin{eqnarray}
U&=&
\left(D'
+\frac{r^2}{2}\sin^2\theta\frac{d\Omega_{\rm F}^2}{d \Psi}\Psi'\right)\Psi', \\
W&=&
\left(
\partial_\theta D
+\frac{r^2}{2}\sin^2\theta\frac{d\Omega_{\rm F}^2}{d \Psi}
\partial_\theta \Psi\right)
\partial_\theta \Psi
+8\pi^2r^2\frac{dI^2}{d\Psi},
\end{eqnarray}
where $I=I(\Psi)$ and $\Psi$ are the electric current and 
the magnetic flux through an axisymmetric polar cap, which are defined by 
Eqs.~(\ref{eq:I-def}) and (\ref{eq:Psi-def}), respectively, and $D$ is defined by
\begin{equation}
D:=\alpha^2-r^2\Omega_{\rm F}^2\sin^2\theta. \label{eq:D-def}
\end{equation}
The derivation of the GS equation 
(\ref{eq:GS-eq}) is given in Appendices \ref{Derivation} and \ref{I-Psi}. 

In general, the GS equation for a black hole magnetosphere  
has three regular singular points: one is at the event horizon 
$\Delta=0$, and the other two are at the light surfaces 
defined by $D=0$. 
A light surface is a timelike hypersurface on which the rotational speed of 
magnetic field lines are equal to the speed of light. 
The inner light surface  given by a function $r=r_{\rm LS-}(\theta)$
has a spacelike section with spherical topology, whereas the 
outer one, $r=r_{\rm LS+}(\theta)$, has that of cylindrical topology. 

In the case of the Kerr spacetime, the Kerr-Schild coordinate system is often adopted,
for example, in numerical simulations (e.g.,~\cite{McKinney:2004,Komissarov:2001,Nagataki:2009}),
since there is no coordinate singularity on the event horizon. 
Therefore, we give the Kerr-Schild coordinate system for the 
RN spacetime in Appendix \ref{KS}. 
In Appendix \ref{KS}, we show that the singular point on the event horizon 
in the GS equation appears even if we adopt Kerr-Schild coordinates.
This is also true in the case of the Kerr spacetime, though 
it is not shown in this paper (we will show it elsewhere). 
As long as a stationary magnetic field is considered, the singular point 
of the basic equation will appear on 
the event horizon, since there is no timelike Killing vector field on or inside the 
event horizon, or in other words, the stationary configuration cannot be realized inside 
the black hole. 

\section{Regularity Conditions and Boundary conditions}

In this section, we consider only the case of $\Omega_{\rm F}\neq0$. 
Then, by virtue of the symmetry of the background spacetime, 
without loss of generality, we may assume $\Omega_{\rm F}>0$. 
The case of $\Omega_{\rm F}=0$ will be treated as 
specific cases later.

\subsection{Symmetry Axis $\theta=0$}

On the symmetry axis $\theta=0$, both $\Psi$ and $I$ should vanish by their definitions. 
$I$ is a function of $\Psi$, and thus, $I|_{\Psi=0}$ should vanish. 

\subsection{Event Horizon}
In Appendix \ref{KS}, we give the components of $F_{ab}$ in the Kerr-Schild 
coordinate system $(T,\Phi,R,\Theta)$. {}From Eq.~(\ref{eq:RT-comp}), we have 
\begin{equation}
F_{R\Theta}=-\frac{\cal M}{\Delta},
\end{equation}
where 
\begin{equation}
{\cal M}=\frac{1}{2\pi}\left(2Mr-r_+^2\right)\Omega_{\rm F}
\partial_\theta\Psi+\frac{2r^2}{\sin\theta}I(\Psi). 
\end{equation}
Since the Kerr-Schild coordinate system is non-singular on the event horizon, 
$F_{R\Theta}$ must be finite there. Thus, we have
${\cal M}|_{r=r_{+}}=0$.
This leads to
\begin{equation}
I+\frac{1}{4\pi r^2}
(2Mr-r_+^2)
\Omega_{\rm F}\sin\theta\partial_\theta\Psi=0~~~~~{\rm at}~~r=r_+~. 
\label{eq:h-regularity}
\end{equation}
The above condition corresponds to the horizon boundary condition derived 
by Znajek for the Kerr black hole~\cite{Znajek:1977}. 
It is seen from the GS equation (\ref{eq:GS-eq}) that, 
in order that $\Psi$, $\Psi'$ and $\Psi''$ are 
finite on the event horizon $r=r_+$, 
the following condition should be satisfied
\begin{equation}
L_\theta\Psi+\frac{W}{D}=0. \label{Horizon}
\end{equation}
However, this condition is satisfied if $\Psi$ satisfies 
the regularity condition (\ref{eq:h-regularity}), 
and thus no additional constraint is imposed by this equation. 

In the extreme case, since the equation $\Delta=0$ has a double root $r_\pm=r_{\rm H}$, 
${\cal M}'|_{r=r_{\rm H}}$ should vanish as well as 
${\cal M}|_{r=r_{\rm H}}=0$ so that $F_{R\Theta}$ is finite 
on the event horizon. These conditions imply  
\begin{eqnarray}
I+\frac{\Omega_{\rm F}}{4\pi}\sin\theta\partial_\theta\Psi&=&0, 
\label{eq:h-regularity-1}\\
\frac{dI}{d\Psi}\Psi'+
\frac{\sin\theta}{4\pi}
\left(\Omega_{\rm F}\partial_\theta\Psi'+\frac{d\Omega_{\rm F}}{d\Psi}\Psi'
\partial_\theta\Psi\right)&=&0,
\label{eq:h-regularity-2}
\end{eqnarray}
on the event horizon $r=r_{\rm H}$. 
We can see from the GS equation (\ref{eq:GS-eq}) that, 
in order that $\Psi$, $\Psi'$, and $\Psi''$ are 
finite on the event horizon for the extreme case, 
not only Eq.~(\ref{Horizon}) but also 
the following condition should be satisfied,  
\begin{equation}
\left(L_\theta\Psi+\frac{W}{D}\right)'=0~~~~{\rm at}~~r=r_{\rm H}. 
\label{Horizon-dash}
\end{equation}
This condition is satisfied if $\Psi$ satisfies 
the regularity conditions (\ref{eq:h-regularity-1}) and (\ref{eq:h-regularity-2}), 
and thus no additional constraint is imposed by this equation. 

\subsection{Light Surface}

As mentioned, the light surfaces are singular points of the GS 
equation (\ref{eq:GS-eq}). 
In the extreme case, the radial coordinates of the light surfaces,  
which are the roots of the equation $D=0$ in the domain $r>r_{\rm H}$, are given by  
\begin{equation}
r=r_{{\rm LS}\pm}=\frac{1\pm\sqrt{1-4r_{\rm H}
\Omega_{\rm F}\sin\theta}}{2\Omega_{\rm F}\sin\theta}.
\label{rLS}
\end{equation}
In order that $\Psi''$ and $L_\theta\Psi$ are finite 
on the light surfaces, $U+W/\Delta$ must vanish there. 
This requirement leads to the following regularity conditions on the light surfaces: 
\begin{equation}
V^a\partial_a\Psi=-8\pi^2r^2\frac{dI^2}{d\Psi}~~~~~~{\rm at}~~r=r_{{\rm LS}\pm}
\label{eq:Neumann}
\end{equation}
where
\begin{equation}
V^a:=r^2g^{ab}\left(\partial_b D+\frac{r^2}{2}\sin^2\theta \partial_b\Omega_{\rm F}^2\right).
\end{equation}

\begin{figure}
\centering
\includegraphics[width=0.80\textwidth]{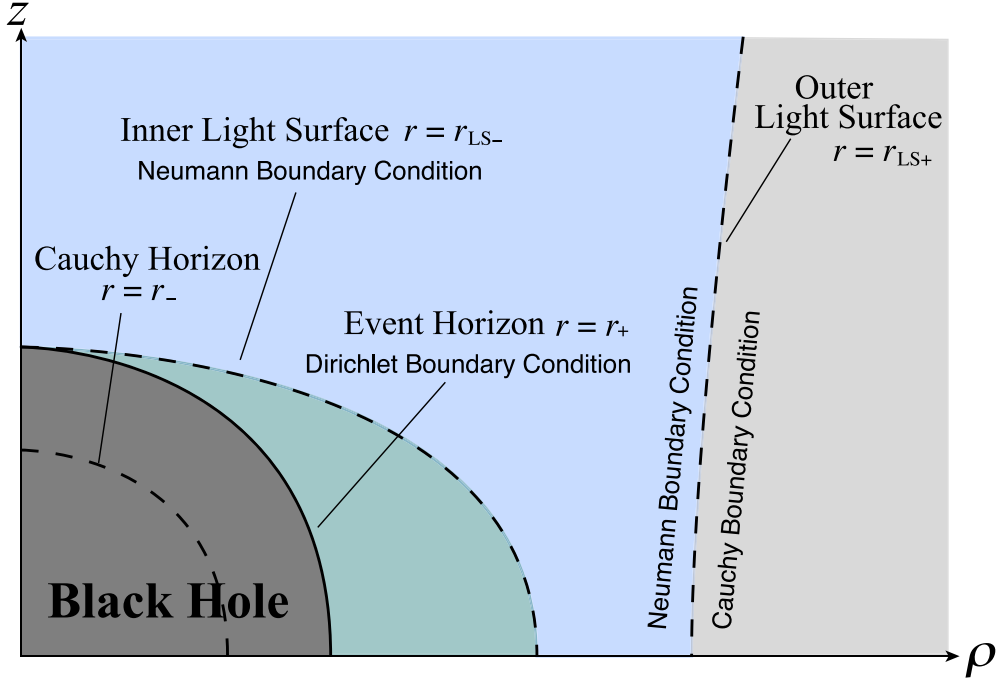}
\caption{
Schematic diagram of the magnetosphere of a non-extreme RN black hole
$r_{+}>r_{-}$, 
where $\rho=r\sin\theta$ and $z=r\cos\theta$. 
}
\label{fg:non-extreme}
\end{figure}

\subsection{Boundary Conditions in the Case of Non-Degenerate Horizons}

For simplicity, in this subsection, we assume that $\Omega_{\rm F}$ is constant.
By virtue of this assumption, the conditions (\ref{eq:Neumann}) 
become Neumann boundary conditions on the light surfaces. 

Here, we consider the non-extreme case $r_{+}>r_{-}$. 
Let us assume that the functional form of $I(\Psi)$ has already been determined 
before solving the GS equation. Then, by solving 
the horizon regularity condition (\ref{eq:h-regularity}), we 
obtain $\Psi$ on the event horizon. 
Since $\Psi$ is the magnetic flux 
through the polar cap (see Appendix B), $\Psi$ should vanish at $\theta=0$. 
Thus, there seems to be no freedom for setting a boundary condition for $\Psi$ in 
solving Eq.~(\ref{eq:h-regularity}), but this is not true. Since $\theta=0$ is a 
regular singular point of Eq.~(\ref{eq:h-regularity}), there still remains one degree of 
freedom for choosing a boundary value of the second-order derivative of $\Psi$. 
Hence, a Dirichlet boundary condition for the GS equation 
is determined on the event horizon by the regularity condition
(\ref{eq:h-regularity}).
By imposing this Dirichlet boundary condition at $r=r_+$ and further 
Neumann boundary conditions at $r=r_{{\rm LS}-}$ (\ref{eq:Neumann}) 
and on the equatorial plane 
(e.g., the reflection-symmetric boundary condition $\partial_\theta\Psi|_{\theta=\pi/2}=0$), 
a solution for the GS equation (\ref{eq:GS-eq}) is uniquely determined  
in the domain $r_+ < r < r_{{\rm LS}-}$. 
By imposing two Neumann boundary conditions (\ref{eq:Neumann}) on the two 
light surfaces $r=r_{{\rm LS}\pm}$ and the equatorial plane $\theta=\pi/2$, and a
further Dirichlet boundary condition $\Psi=0$ on the symmetry axis $\theta=0$, 
a solution for the GS equation is uniquely determined 
in the domain $r_{{\rm LS}-}< r < r_{{\rm LS}+}$. 
For the domain $r_{{\rm LS}+} < r <\infty$, 
if we impose a boundary condition at $\Psi$ for $r\rightarrow\infty$, 
then we can obtain a solution 
for the GS equation. 

The GS equation can be solved for these three domains, 
$r_{+} < r < r_{{\rm LS}-}$, $r_{{\rm LS}-}< r < r_{{\rm LS}+}$, and 
$r_{{\rm LS}+} < r < \infty$,
independently by the above procedure (see Fig.~\ref{fg:non-extreme}). 
Thus, for an arbitrary electric current $I(\Psi)$, 
the obtained solution for $\Psi$ is, in general, 
not $C^1$ but at most $C^{1-}$ at the boundaries $r=r_{{\rm LS}\pm}$. 
It was first reported by 
Contopoulos, Kazanas, and Fendt (CKF) 
that the continuity of the 
first-order derivative of $\Psi$ at the light surface as well as the 
continuity of $\Psi$ itself determines the 
functional form of the electric current $I(\Psi)$ in the case of the 
pulsar magnetosphere~\cite{Contopoulos:1999}. 
They numerically obtained $C^1$ solutions for $\Psi$ 
by an iterative method in which both $\Psi$ and the functional form of 
$I(\Psi)$ are determined simultaneously. 
The CKF method was used by several authors to study the pulsar magnetosphere and 
they showed that this method also suitable for their case of interest~\cite{Ogura:2003,Gruzinov:2005,Timokhin:2006}.
However, it should be noted that a mathematically
rigorous proof for the existence of the $C^1$ solution for the GS equation 
has not yet been given. 

Although there is at most one light surface in the case of the 
pulsar magnetosphere, there can be two light surfaces if $\Omega_{\rm F}$ is 
a non-vanishing constant in the case of the black hole magnetosphere  
(see Fig.~\ref{fg:non-extreme}). 
Thus, if we chose the functional form of $I(\Psi)$ 
such that $\Psi$ is $C^1$ in the domain $r_{+} < r < r_{{\rm LS}+}$,  
imposing an asymptotic boundary condition for $r\rightarrow\infty$ 
does not guarantee the continuity of the derivative of $\Psi$ at the outer light surface $r=r_{{\rm LS}+}$.
Thus, in order to obtain a solution which is $C^1$ in the domain 
$r_+<r<\infty$, we need to solve the GS equation 
for the outermost domain $r_{{\rm LS}+}< r <\infty$ as a Cauchy problem with 
a boundary data for $\Psi$ and the derivatives of $\Psi$ on $r=r_{{\rm LS}+}$. 
This implies that we cannot impose the asymptotic boundary condition for 
$r\rightarrow \infty$. 
In general, it is difficult to solve an elliptic-type differential 
equation numerically, such as the GS equation, as a Cauchy problem due to the numerical instability. 
Thus, in the case with two light surfaces, it is difficult to 
numerically obtain a solution for the GS equation in 
the outermost domain $r_{{\rm LS}+}<r<\infty$. 
However, this might not be a serious problem, since we may understand whether the 
Blandford-Znajek mechanism works by studying only 
the domain $r_+<r<r_{{\rm LS}+}$. 

Uzdensky applied the CKF method to the magnetospheres of the 
Schwarzschild black hole~\cite{Uzdensky:2004} and of the  
Kerr black hole~\cite{Uzdensky:2005}, 
but he focused on only the cases in which 
there is only one light surface   
by virtue of a particular assumption on $\Omega_{\rm F}$: 
Uzdensky assumed that $\Omega_{\rm F}$ asymptotically decreases,  
and hence there is no outer light surface. 
Thus, Uzdensky succeeded 
in numerically obtaining global solutions without solving Cauchy problems for the 
GS equation. 

\subsection{Boundary Conditions in the Case of a Degenerate Horizon}

\begin{figure}
\centering
\includegraphics[width=0.8\textwidth]{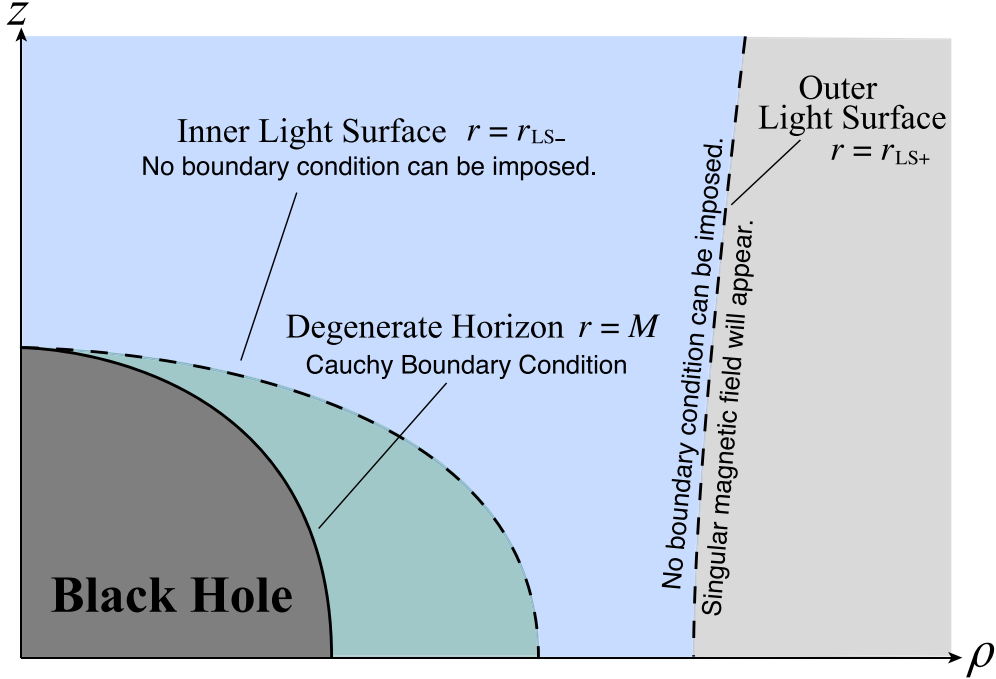}
\caption{
The same as Fig.~\ref{fg:non-extreme}, 
but for an extreme RN black hole $r_{+}=r_{-}=r_{\rm H}$. 
}
\label{fg:extreme}
\end{figure}

Here, we consider the extreme case $r_{+}=r_{-}=r_{\rm H}$, 
which is the main case of interest in this paper. 
In this subsection, we also assume that $\Omega_{\rm F}$ is constant. 
The horizon regularity conditions 
(\ref{eq:h-regularity-1}) and (\ref{eq:h-regularity-2}) give boundary values of 
$\Psi$ and the derivative of $\Psi$. Thus, in the extreme case, we must 
solve the GS equation (\ref{eq:GS-eq}) as a Cauchy problem even for the 
domain $r_{\rm H}<r<r_{{\rm LS}-}$ (see Fig.~\ref{fg:extreme}). 
As mentioned, it is difficult to numerically solve the GS equation 
as a Cauchy problem, and hence, it seems to be difficult to numerically
obtain a solution for the physically important 
domain $r_{\rm H}<r<r_{{\rm LS}+}$ in the extreme case. 
Further, even if we find a procedure for numerically solving the GS equation as a Cauchy problem, the regularity condition 
on the inner light surface $r=r_{{\rm LS}-}$ may not be satisfied for an arbitrary
functional form of electric current $I(\Psi)$: we must assume the functional 
form of $I(\Psi)$ to solve the GS equation as a Cauchy problem, but, in general, 
the assumed electric current $I(\Psi)$ does not satisfy the 
regularity condition on the inner light surface. 
As a result, it seems to be impossible to obtain a solution for the GS equation numerically, 
which is finite on the inner light surface, in the extreme case. 
In this sense, the perturbative analytic approach discussed in \S VI
is very important.

We should note that 
even if we find analytically the electric current $I$ which guarantees 
the finiteness of $\Psi$ 
and its derivative on the inner light surface, such a electric current $I$ might not 
guarantee the finiteness of both $\Psi$ and its derivative  
on the outer light surface. 
This implies that either the force-free condition should break down near the outer light surface 
or the rotational velocity should decay far from the black hole so that the outer light 
surface does not exist, as in the situation studied by Uzdensky. 

\section{Vacuum case}

In this section, we consider the vacuum case $I=0$. 

\subsection{$\Omega_{\rm F}=0$ on the horizon}

Here, we consider the case of $\Omega_{\rm F}=0$ on the event horizon, $r=r_{+}$. 
Even if $\Omega_{\rm F}$ does not vanish except on the event horizon, it satisfies
\begin{equation}
\partial_\theta\Omega_{\rm F}=\frac{d\Omega_{\rm F}}{d\Psi}\partial_\theta\Psi=0
~~~~~{\rm at}~~r=r_{+}.
\label{eq:del-Omega}
\end{equation}
Thus, on the event horizon, $\partial_\theta\Psi=0$ or $d\Omega_{\rm F}/d\Psi=0$ 
should hold. 
In the former case, $\Psi$ vanishes on the event horizon, implying that 
the magnetic flux does not penetrate the event horizon. 
In the latter case, from Eq.~(\ref{eq:GS-eq}), in order that $\Psi$, $\Psi'$, and $\Psi''$ 
are finite on the event horizon, the following equation should be satisfied:
\begin{equation}
L_\theta\Psi+r^2\Psi'\left(\frac{\Delta}{r^2}\right)'=0~~~~~~{\rm at}~~r=r_+ . 
\label{eq:h-regularity-0}
\end{equation}
In the extreme case, since $\Delta'=0$ also holds  
on the event horizon $r=r_{\rm H}$, we have 
\begin{equation}
L_\theta \Psi =0~~~~~{\rm at}~~r=r_{\rm H}.
\end{equation}
The solution of the above equation which satisfies the regularity 
condition on the symmetry axis $\theta=0$ is
\begin{equation}
\Psi=C(1-\cos\theta), \label{eq:monopole}
\end{equation}
where $C$ is an integration constant. The above solution implies that the magnetic field 
which can penetrate the event horizon is the only monopole component. 
By contrast to the extreme case, non-monopole components can penetrate 
the event horizon in the non-extreme case, 
since, in this case, Eq.~(\ref{eq:h-regularity-0}) does not necessarily 
imply $L_\theta\Psi=0$. 
As a result, we can conclude that the Meissner-like effect of the extreme 
black hole appears in the vacuum case. 

In the case that $\Omega_{\rm F}$ vanishes everywhere, we can obtain global solutions. 
The solutions which satisfy the regularity condition on the symmetry axis $\theta=0$ 
are written in the form
\begin{equation}
\Psi=\sum_{l=0}^\infty R_l(r)P_l^1(\cos\theta)\sin\theta,
\label{eq:non-monopole}
\end{equation}
where $P_l^1(x)$ is the  associated Legendre function of the first kind with $m=1$. 
Then, the GS equation (\ref{eq:GS-eq}) becomes
\begin{equation}
\alpha^2 R_l''+2\alpha'\alpha R_l'-\frac{l(l+1)}{r^2}R_l=0.
\label{eq:Rl}
\end{equation}
In the extreme case, in order that 
$R_l$, $R_l'$, and $R_l''$ are finite on the event horizon, 
$R_l$ of $l\geq 1$ vanishes, and the only monopole component $l=0$ may remain. 

\subsection{$\Omega_{\rm F}\neq0$ on the horizon}

In this case, from the regularity condition on the event horizon 
(\ref{eq:h-regularity}), we have
\begin{equation}
\partial_\theta\Psi=0,
\end{equation}
for both extreme and non-extreme cases. 
Thus, $\Psi=0$ is a solution which satisfies the regularity condition on the 
symmetry axis $\theta=0$. The magnetic field does not penetrate the 
event horizon at all. 

\section{Case with an electric current}

In this section, we focus on the extreme case $r_\pm=r_{\rm H}$ 
and assume that the rotational velocity of the magnetic field 
$\Omega_{\rm F}$ is constant. 

In the case of $\Omega_{\rm F}=0$, it is seen from Eq.~(\ref{eq:GS-eq}) that since 
$D=\alpha^2$, the following condition should be satisfied on the event horizon: 
\begin{equation}
\frac{dI^2}{d\Psi}=0.
\end{equation}
The above condition allows $I={\rm const.}$ on the event horizon. 
However, since $I$ should vanish on the symmetry axis $\theta=0$, the allowed constant is zero.
Thus, in this case, the same argument as was used in the vacuum case discussed in the previous section is also true. The allowed 
configuration of the magnetic field on the horizon is only the 
monopole component (\ref{eq:monopole}). Hence, hereafter, we focus on the case of 
$\Omega_{\rm F}\neq0$. 

\subsection{Grad-Shafranov equation near the event horizon}
We are interested in 
the configuration of the magnetic field near the event horizon.
In order to analyze the GS equation, we introduce the 
following dimensionless quantities:
\begin{eqnarray}
r&=:&r_{\rm H} y, \\
\Psi&=:&r_{\rm H} \psi, \\
\varepsilon&:=&r_{\rm H} \Omega_{\rm F}, \\
I(\Psi)&=:&\varepsilon {\cal I}(\psi), \label{calI-def}\\
8\pi^2\frac{dI^2}{d\Psi}&=:&\varepsilon^2r_{\rm H}^{-1}{\cal S}(\psi). \label{calS-def}
\end{eqnarray}
Using these quantities, the GS equation (\ref{eq:GS-eq}) becomes 
\begin{equation}
\partial_y^2\psi+\frac{1}{(y-1)^2}
L_\theta\psi 
+\frac{{\cal U}}{\cal D}
+\frac{{\cal W}}{(y-1)^2{\cal D}}=0,
\label{GS}
\end{equation}
where
\begin{eqnarray}
{\cal D}&=&y\left[(y-1)^2-\varepsilon^2 y^4\sin^2\theta\right], \\
{\cal U}&=&2(y-1-\varepsilon^2y^4\sin^2\theta)\partial_y\psi, \\
{\cal W}&=&-\varepsilon^2 y^5\left[\sin2\theta \partial_\theta\psi
-{\cal S}\left(\psi\right)\right].
\end{eqnarray}
The regularity conditions on the event horizon 
(\ref{eq:h-regularity-1}) and (\ref{eq:h-regularity-2}) become
\begin{eqnarray}
{\cal I}(\psi)+\frac{1}{4\pi}\sin\theta\partial_\theta\psi&=&0, 
\label{HBo} \\
\frac{d{\cal I}}{d\psi}\partial_y\psi+
\frac{1}{4\pi}\sin\theta\partial_\theta(\partial_y\psi)&=&0.
\label{HBDo}
\end{eqnarray}
The regularity conditions on the light surfaces (\ref{eq:Neumann}) become
\begin{equation}
\partial_\theta\psi-\frac{{\cal S}(\psi)}{\sin2\theta}
-\varepsilon y \tan\theta \sin\theta (1-\varepsilon y^2\sin\theta)\partial_y\psi=0
~~~~~{\rm at}~~
y=y_{{\rm LS}\pm}, 
\label{LS}
\end{equation}
where 
\begin{equation}
y_{{\rm LS}\pm}:=\frac{r_{{\rm LS}\pm}}{r_{\rm H}}.
\end{equation}
Here, we assume that $\psi$ can be written in the form of Taylor series around 
the event horizon, $y=1$, as
\begin{equation}
\psi=\sum_{n=0}^\infty\psi^{(n)}(\theta)(y-1)^n.
\label{Taylor}
\end{equation}
The coefficients $\psi^{(n)}$ are, in principle, determined by 
the GS equation (\ref{GS}) with the regularity conditions 
(\ref{HBo}), (\ref{HBDo}), and (\ref{LS}). 
Using the expression (\ref{Taylor}), 
Eqs.~(\ref{HBo}) and (\ref{HBDo}) can be rewritten in the forms
\begin{eqnarray}
{\cal I}(\psi^{(0)})+\frac{1}{4\pi}\sin\theta\frac{d\psi^{(0)}}{d\theta}&=&0, 
\label{HB} \\
\frac{d{\cal I}}{d\psi}(\psi^{(0)})\psi^{(1)}+
\frac{1}{4\pi}\sin\theta\frac{d\psi^{(1)}}{d\theta}&=&0. 
\label{HBD} 
\end{eqnarray}
If we fix the functional form of ${\cal I}(\psi)$, 
we obtain $\psi^{(0)}$ and $\psi^{(1)}$ from the above equations, 
and further, we obtain $\psi^{(n)}$ of $n\geq2$ from Eq.~(\ref{GS});  
in order to get $\psi^{(n)}$ for $n\geq3$, 
we use an equation obtained by $(n-2)$-times differentiation of Eq.~(\ref{GS}) 
with respect to $y$. 
For example, for $n=2$, by evaluating 
the GS equation (\ref{GS}) on the event horizon $y=1$, we have
\begin{eqnarray}
&&\varepsilon^2\left[L_\theta\psi^{(2)}+2\psi^{(2)}
+2\cot\theta\left(\frac{d\psi^{(2)}}{d\theta}
-\frac{1}{\sin2\theta}\frac{d{\cal S}}{d\psi}(\psi^{(0)})\psi^{(2)}
\right) \right]
\nonumber \\
&&=-2\cot\theta
\left(\frac{d\psi^{(0)}}{d\theta}-\frac{{\cal S}(\psi^{(0)})}{\sin2\theta}\right)
\left(\frac{1}{\sin^2\theta}+12\varepsilon^2\right)
-\varepsilon^2\psi^{(1)}\left[2
-\frac{1}{2\sin^2\theta}\frac{d^2{\cal S}}{d\psi^2}(\psi^{(0)}){\psi^{(1)}}\right].
\label{GS-2} \nonumber \\
\end{eqnarray}
Here, we should again note that ${\cal I}(\psi)$ cannot be freely specified. 
The functional form of ${\cal I}(\psi)$ must be chosen such that the 
regularity condition (\ref{LS}) on the inner light surface is satisfied. 

\subsection{Perturbative analysis}

We consider slowly rotating magnetic fields, or in other words, 
we assume $0<\varepsilon \ll 1$. We rewrite the basic equations  
in the form of the power series with respect to $\varepsilon$, and then 
we construct a solution of $\psi$ on the horizon, i.e., $\psi^{(0)}$, by 
perturbative procedures with respect to $\varepsilon$. 
Although, as mentioned, it seems to be impossible to 
determine the functional form of $\cal I$ numerically, 
we can find it by this method. 

In order to construct a perturbative solution for $\psi^{(n)}$, we write 
\begin{equation}
\psi^{(n)}(\theta)=\sum_{N=0}^\infty\psi^{(n)}_N\varepsilon^N.
\end{equation}
Further, we assume 
\begin{equation}
{\cal I}(x)=\sum_{N=0}^\infty{\cal I}_{N+1}(x)\varepsilon^N
~~~~~{\rm and}~~~~~
{\cal S}(x)=\sum_{N=0}^\infty{\cal S}_{N+2}(x)\varepsilon^N.
\label{assumptions}
\end{equation}
{}From Eq.~(\ref{rLS}), the location of the inner light surface is written as
\begin{equation}
y_{{\rm LS}-}=1+\sum_{n=1}^\infty\frac{(2n)!}{n!(n+1)!}(\varepsilon\sin\theta)^n
=1+\varepsilon\sin\theta +2(\varepsilon\sin\theta)^2
+\cdots. 
\end{equation}
Because $y_{\rm LS-}-1={\cal O}(\varepsilon)$, we can express the quantities 
on the inner light surface by using the quantities on the event horizon. 
For example, $\partial_\theta\psi$ at $y=y_{{\rm LS}-}$ is written as
\begin{equation}
 \partial_\theta\psi(y_{{\rm LS}-},\theta)
 =\frac{d\psi^{(0)}_0}{d\theta}+\varepsilon\left(\frac{d\psi^{(0)}_1}{d\theta}
 +\sin\theta\frac{d\psi^{(1)}_0}{d\theta}\right)+\cdots.
\end{equation}

Since the main purpose of this study is to see the effect of an electric current
on the configuration of the magnetic field on the event horizon,
we focus on $\psi^{(0)}$.
For this purpose, we rewrite Eqs.~(\ref{HBD}) and (\ref{GS-2}) in more 
appropriate forms as follows. 

By differentiating Eq.~(\ref{HB}) with respect to $\theta$, we have
\begin{equation}
 4\pi\frac{d{\cal I}}{d\psi}(\psi^{(0)})
 +\left(\frac{d\psi^{(0)}}{d\theta}\right)^{-1}
 \frac{d}{d\theta}\left(\sin\theta\frac{d\psi^{(0)}}{d\theta}\right)=0.
 \label{HB-diff}
\end{equation}
{}From Eq.~(\ref{HBD}), we have
\begin{equation}
 4\pi\frac{d{\cal I}}{d\psi}(\psi^{(0)})
 +\sin\theta\frac{d}{d\theta}\ln\psi^{(1)}=0.
\end{equation}
By subtracting Eq.~(\ref{HB-diff}) from the above equation, we obtain
\begin{equation}
 \left(\frac{d\psi^{(0)}}{d\theta}\right)^{-1}\frac{d}{d\theta}
\left(\sin\theta\frac{d\psi^{(0)}}{d\theta}\right)
 -\sin\theta\frac{d}{d\theta}\ln\psi^{(1)}=0.
\end{equation}
It is easy to integrate the above equation, and we have 
\begin{equation}
  \psi^{(1)}=C'\sin\theta\frac{d\psi^{(0)}}{d\theta},
 \label{eq:psi1}
\end{equation}
where $C'$ is an integration constant. In order to obtain $\psi^{(1)}$, we use 
Eq.~(\ref{eq:psi1}) rather than Eq.~(\ref{HBD}). 

By substituting Eq.~(\ref{HB}) into Eq.~(\ref{HB-diff}) 
and using Eqs.~(\ref{calI-def}) and (\ref{calS-def}), we have
\begin{equation}
 -2\cot\theta\left(\frac{d\psi^{(0)}}{d\theta}
-\frac{{\cal S}(\psi^{(0)})}{\sin2\theta}\right)=L_\theta\psi^{(0)}.
 \label{Horizon-2}
\end{equation}
The above equation is equivalent to Eq.~(\ref{Horizon}). 
By substituting Eq.~(\ref{Horizon-2}) into the first term on the right-hand side of 
Eq.~(\ref{GS-2}), we obtain
\begin{eqnarray}
 L_\theta\psi^{(0)}
&=&\varepsilon^{2}\sin^2\theta\Bigg{[}-12L_\theta\psi^{(0)}+L_\theta\psi^{(2)}+2\psi^{(2)}
 \nonumber \\
&& \left.+2\cot\theta
   \left(\frac{d\psi^{(2)}}{d\theta}-\frac{1}{\sin2\theta}
   \frac{d{\cal S}}{d\psi}\psi^{(2)}\right)
   +\psi^{(1)}\left(2-\frac{1}{2\sin^2\theta}\frac{d^2{\cal S}}{d\psi^2}
   \psi^{(1)}\right)\right].
\label{eq:psi0}
\end{eqnarray}
We shall use the above equation rather than Eq.~(\ref{GS-2}).

\subsubsection{Zeroth-Order Solutions for $\psi^{(0)}$}

Here, we obtain the zeroth-order solutions for $\psi^{(0)}$.
Hereafter, the arguments of ${\cal S}_N$ and ${\cal I}_N$ are $\psi^{(0)}_0$ 
as long as we do not specify them.

First of all, we write down 
the equations to obtain the zeroth-order solutions for $\psi^{(0)}$.
{}From the lowest order of Eqs.~(\ref{LS}), (\ref{HB}), and (\ref{eq:psi0}),
we have 
\begin{eqnarray}
 \frac{d\psi^{(0)}_0}{d\theta}-\frac{{\cal S}_2}{\sin2\theta} &=&0,
 \label{LS00}\\
 {\cal I}_1+\frac{1}{4\pi}\sin\theta\frac{d\psi^{(0)}_0}{d\theta}&=&0, 
 \label{HB0} \\
 L_\theta\psi^{(0)}_0&=&0,
 \label{eq:psi00}
\end{eqnarray}
where from Eqs.~(\ref{calI-def}), (\ref{calS-def}), and (\ref{assumptions}), 
\begin{equation}
 {\cal S}_2=16\pi^2{\cal I}_1\frac{d{\cal I}_1}{d\psi}. \label{S2}
\end{equation}
We can easily integrate Eq.~(\ref{eq:psi00}) and obtain
\begin{equation}
 \psi^{(0)}_0=C^{(0)}_0(1-\cos\theta),
 \label{00-sol}
\end{equation}
where $C^{(0)}_0$ is an integration constant. 
The above result implies that $\psi^{(0)}_0$ has only the monopole component.
Then, substituting Eq.~(\ref{00-sol}) into Eq.~(\ref{HB0}),
we have  
\begin{equation}
{\cal I}_1(X)=-\frac{1}{4\pi}C^{(0)}_0\hat{X}
\left(2-{\hat X}\right), 
\label{I1}
\end{equation}
where 
\begin{equation}
{\hat X}=\frac{X}{C^{(0)}_0}. \label{Xhat}
\end{equation}

It is non-trivial whether the lowest order of the inner 
light surface regularity condition 
(\ref{LS00}) is satisfied by $\psi^{(0)}_0$ and ${\cal I}_1$ obtained above. 
{}From Eqs.~(\ref{I1}) and (\ref{S2}), we have
\begin{equation}
 {\cal S}_2(X)=16\pi^2{\cal I}_1(X)\frac{d{\cal I}_1}{dX}(X)
 =2C^{(0)}_0{\hat X}\left(2-3{\hat X}+{\hat X}^2\right).
 \label{S0-sol}
\end{equation}
It is easy to check that Eqs.~(\ref{00-sol}) and (\ref{S0-sol}) satisfy Eq.~(\ref{LS00}).
Namely, we have obtained a small electric current which satisfies the lowest order of the 
inner light surface regularity condition.

It is worthwhile to notice the meaning of the zeroth-order solutions 
for $\psi^{(0)}$, i.e., $\psi^{(0)}_0$.
In the limit $\varepsilon \rightarrow 0$, $\Omega_{\rm F}$ and $I$ become zero, whereas 
$\psi^{(0)}$ becomes $\psi^{(0)}_0$.
Since the case $\Omega_{\rm F}=I=0$ corresponds to the vacuum case, i.e., 
the case without an electric current, $\psi^{(0)}_0$ corresponds to the vacuum solution. 
As we showed in \S V, the vacuum solution 
has only the monopole component on the event horizon.
Eq.~(\ref{00-sol}) is consistent with this fact. 
The small electric current $I=\varepsilon{\cal I}_1$ can be regarded as 
a result of the slowly rotating monopole field $\psi^{(0)}_0$. 

\subsubsection{First-Order Solutions for $\psi^{(0)}$}

Next, we consider the correction of $O(\varepsilon^1)$ 
to the zeroth-order solution for $\psi^{(0)}$.
Hereafter, we assume $C^{(0)}_0\neq 0$. In our perturbative method, 
the case of $C^{(0)}_0=0$ is quite 
different from the case $C^{(0)}_0\neq0$.
If we choose $C^{(0)}_0=0$, we can obtain only the trivial solution $\psi^{(0)}=0$ 
using our perturbative method. We prove this statement 
in Appendix {\ref{Proof}}.

The equations determining $\psi^{(0)}_1$ are derived from
Eqs.~(\ref{LS}), (\ref{HB}), and (\ref{eq:psi0}) of $O(\varepsilon^1)$:
\begin{eqnarray}
 &&\frac{d\psi^{(0)}_1}{d\theta}-\frac{1}{\sin2\theta}
 \left(\frac{d{\cal S}_2}{d\psi}\psi^{(0)}_1+{\cal S}_3\right)
 \nonumber \\
 &&\ \ \ \ \ \ \ \ \ +\sin\theta
 \left(\frac{d\psi^{(1)}_0}{d\theta}
-\frac{1}{\sin2\theta}\frac{d{\cal S}_2}{d\psi}\psi^{(1)}_0
 -\tan\theta\psi^{(1)}_0\right)=0, 
 \label{LS1} \\ 
 &&\frac{d{\cal I}_1}{d\psi}\psi^{(0)}_1+{\cal I}_2 
 +\frac{1}{4\pi}\sin\theta\frac{d\psi^{(0)}_1}{d\theta}\ =\ 0, 
 \label{HB1} \\
 &&L_\theta\psi^{(0)}_1\ =\ 0,  
 \label{eq:psi01}
\end{eqnarray}
where
\begin{equation}
 {\cal S}_3=16\pi^2\left({\cal I}_1\frac{d{\cal I}_2}{d\psi}
+{\cal I}_2\frac{d{\cal I}_1}{d\psi}\right).
 \label{S3}
\end{equation}
We can see from Eq.~(\ref{eq:psi01}) that $\psi^{(0)}_1$ is also the 
monopole solution. 
Thus, the first-order correction merely adds a constant of 
$O(\varepsilon^1)$ to the 
integration constant of the zeroth-order solution.
As a result, without loss of generality, we may assume for the first-order solutions that
\begin{equation}
\psi^{(0)}_1=0.
\end{equation}
{}From Eqs.~(\ref{HB1}), (\ref{S3}), and the above equation, we have 
\begin{equation}
 {\cal I}_2=0={\cal S}_3.
\end{equation}
We should check the inner light surface regularity condition (\ref{LS}).
Since $\psi^{(0)}_1={\cal I}_2={\cal S}_3=0$, Eq.~(\ref{LS1}) becomes
\begin{equation}
\frac{d\psi^{(1)}_0}{d\theta}-\frac{1}{\sin2\theta}
\frac{d{\cal S}_2}{d\psi}
\psi^{(1)}_0-\tan\theta \psi^{(1)}_0=0.
\label{LS10}
\end{equation}
In order to estimate the above equation, 
we need $\psi^{(1)}_0$, which is the zeroth-order solution for $\psi^{(1)}$. 
We obtain $\psi^{(1)}_0$ from Eq.~(\ref{eq:psi1}) of $O(\varepsilon^0)$ as
\begin{equation}
 \psi^{(1)}_0=C'\sin\theta\frac{d\psi^{(0)}_0}{d\theta},
\end{equation}
where we assume that $C'$ is the order of unity.
Substituting Eq.~(\ref{00-sol}) into the above equation, we obtain 
\begin{equation}
 \psi^{(1)}_0=C'C^{(0)}_0\sin^2{\theta}.
 \label{10-sol}
\end{equation}
By substituting Eq.~(\ref{10-sol}) into Eq.~(\ref{LS10}) 
and using the functional form of ${\cal S}_2$ given by Eq.~(\ref{S0-sol}), 
we can see that Eq.~(\ref{LS10}) is satisfied. Here it is worthwhile to notice that $\psi^{(1)}_0$, as well as $\psi^{(0)}_0$, necessarily 
corresponds to a vacuum solution.
It is easy to check that Eq.~(\ref{10-sol}) is consistent with 
Eq.~(\ref{eq:non-monopole}).

\subsubsection{Second-Order Solutions for $\psi^{(0)}$}

Hereafter, we will make frequent use of Eqs.~(\ref{LS00}) and (\ref{LS10}) without giving an explicit reference.
Now, we consider the correction of $O(\varepsilon^2)$ to the zeroth-order solution for $\psi^{(0)}$.
The equations determining $\psi^{(0)}_2$ 
can be obtained from Eqs.~(\ref{LS}), (\ref{HB}), and (\ref{eq:psi0})
of $O(\varepsilon^2)$:

\begin{eqnarray}
&&\frac{d\psi^{(0)}_2}{d\theta}-\frac{1}{\sin2\theta}
 \left( 
 \frac{d{\cal S}_2}{d\psi}\psi^{(0)}_2+{\cal S}_4
 \right)
 =-\sin^2\theta
 \Biggl[
 \left(2\tan\theta-\frac{1}{2\sin2\theta}
 \frac{d^2{\cal S}_2}{d\psi^2}\psi^{(1)}_0\right) 
 \psi^{(1)}_0 
\nonumber \\
&&~~~~~~~~~~~~~~~~~~~~~~~~~~~~~~~~~~~~~~~~~~+\frac{d\psi^{(2)}_0}{d\theta}
-\frac{1}{\sin2\theta}
\frac{d{\cal S}_2}{d\psi}\psi^{(2)}_0
-2\tan\theta\psi^{(2)}_0
\Biggr],
\label{LS02} \\
&& \frac{d{\cal I}_1}{d\psi}\psi^{(0)}_2+{\cal I}_3 
 +\frac{1}{4\pi}\sin\theta\frac{d\psi^{(0)}_2}{d\theta}=0,
 \label{HB2}\\
&&L_\theta\psi^{(0)}_2
=\sin^2\theta\Bigg{[}L_\theta\psi^{(2)}_0+2\psi^{(2)}_0
+2\cot\theta
   \left(\frac{d\psi^{(2)}_0}{d\theta}
   -\frac{1}{\sin2\theta}\frac{d{\cal S}_2}{d\psi}\psi^{(2)}_0\right) \nonumber \\
&&~~~~~~~~+\psi^{(1)}_0\left(2-\frac{1}{2\sin^2\theta}
   \frac{d^2{\cal S}_2}{d\psi^2}\psi^{(1)}_0\right)\Bigg{]},
\label{eq:psi02} 
\nonumber \\
\end{eqnarray}
where
\begin{equation}
 {\cal S}_4
 =16\pi^2\left(
 {\cal I}_1\frac{d{\cal I}_3}{d\psi}+{\cal I}_3\frac{d{\cal I}_1}{d\psi}
 \right).
\end{equation}
In order to derive the above equations, we have used $\psi^{(0)}_1={\cal I}_2=0$ and 
$\psi^{(1)}_1=0$ obtained by substituting $\psi^{(0)}_1=0$ into Eq.~(\ref{eq:psi1}). 

It should be noted that $\psi^{(2)}_0$ appears in Eqs.~(\ref{LS02}) and (\ref{eq:psi02}). 
In order to determine $\psi^{(2)}_0$, we use Eq.~(\ref{Horizon-2}) 
of $O(\varepsilon^2)$: 
\begin{equation}
 L_\theta\psi^{(0)}_2
 +2\cot\theta\left[\frac{d\psi^{(0)}_2}{d\theta}-\frac{1}{\sin2\theta}
 \left(\frac{d{\cal S}_2}{d\psi}\psi^{(0)}_2+{\cal S}_4\right)\right]=0.
\end{equation}
By substituting Eq.~(\ref{LS02}) into the above equation, we have
\begin{equation}
 L_\theta\psi^{(0)}_2
 =2\cot\theta\sin^2\theta
 \Bigg[\left(2\tan\theta-\frac{1}{2\sin2\theta}
 \frac{d^2{\cal S}_2}{d\psi^2}\psi^{(1)}_0\right)\psi^{(1)}_0
 +\frac{d\psi^{(2)}_0}{d\theta} 
 -\frac{1}{\sin2\theta}
 \frac{d{\cal S}_2}{d\psi}\psi^{(2)}_0-2\tan\theta\psi^{(2)}_0\Bigg].
\end{equation}
By subtracting the above equation from Eq.~(\ref{eq:psi02}), we obtain the 
equation for $\psi^{(2)}_0$ as 
\begin{equation}
 L_\theta\psi^{(2)}_0+6\psi^{(2)}_0-2\psi^{(1)}_0=0.
 \label{20-eq}
\end{equation}
It is easily seen from Eq.~(\ref{10-sol}) 
that $\psi^{(2)}_0=\psi^{(1)}_0/2$ is 
a particular solution for the above equation. 
Thus, the general solution of the above equation is 
expressed by a linear combination of this particular solution and general 
solutions of the following homogeneous equation: 
\begin{equation}
L_\theta f+6f=0.
\label{Homo-eq}
\end{equation}
The general solution of Eq.~(\ref{Homo-eq}) is given by
\begin{equation}
f=\sin\theta\left[c_{\rm p}P_2{}^1(\cos\theta)
+c_{\rm q}Q_2{}^1(\cos\theta)\right],
\end{equation}
where $c_{\rm p}$ and $c_{\rm q}$ are arbitrary constants, and $P_2{}^1$ and 
$Q_2{}^1$ are the associated Legendre functions of  the first and second kinds with 
$l=2$ and $m=1$, respectively. {}From the boundary condition 
at $\theta=0$, $c_{\rm q}$ must vanish. 
Hence, the most general solution of Eq.~(\ref{20-eq}), 
which satisfies the boundary condition 
at $\theta=0$, is given by
\begin{equation}
\psi^{(2)}_0=C^{(2)}_0\sin^2\theta\cos\theta+\frac{C'C^{(0)}_0}{2}\sin^2\theta,
\label{20-sol}
\end{equation}
where we have used $P_2{}^1=(3/2)\sin2\theta$, and $C^{(2)}_0$ is an arbitrary constant.
Here it is worthwhile to notice that the above result also corresponds to a vacuum solution. 
It is easy to check that Eq.~(\ref{20-sol}), as well as $\psi^{(0)}_0$ and $\psi^{(1)}_0$, is consistent with Eq.~(\ref{eq:non-monopole}).

By using Eq.~(\ref{S0-sol}) and substituting 
Eq.~(\ref{10-sol}) into Eq.~(\ref{eq:psi02}), we have   
\begin{equation}
L_\theta\psi^{(0)}_2=
\left[
3C^{(0)}_0C'(1+2C'\cos\theta)-4C^{(2)}_0\cos\theta
\right]\sin^4\theta. 
\label{02-eq}
\end{equation}
The above equation implies that, in general, the magnetic field on the event horizon 
includes non-monopole components of $O(\varepsilon^2)$.

We can easily integrate this equation and obtain 
\begin{equation}
\psi^{(0)}_2=\frac{\sin^2\theta}{60}\left[
2\left(3C^{(0)}{C'}^2-2C^{(2)}_0\right)\cos\theta\left(3\cos^2\theta-7\right)
+15C^{(0)}_0C'\left(\cos^2\theta-5\right)\right],
\label{02-sol}
\end{equation}
where we have chosen the integration constant so that 
$\psi^{(0)}_2|_{\theta=0}=\psi^{(0)}_2|_{\theta=\pi}=0$, i.e.,  
this correction consists of only non-monopole components. 

The functional form of ${\cal I}_3$ is determined by 
using Eq.~(\ref{HB2}) as 
\begin{eqnarray}
&&{\cal I}_3(X)=\frac{1}{120\pi} 
\hat{X}^2(2-\hat{X})^2
\nonumber \\
&&~~~\times \left[
15C^{(0)}_0C'\left(1-\hat{X}\right)
+\left(3C^{(0)}_0{C'}^2-2C^{(2)}_0\right)
\left(2-18\hat{X}+9\hat{X}^2\right)\right],
\end{eqnarray} 
where $\hat{X}$ is defined by Eq.~(\ref{Xhat}).

\subsubsection{Solution near the inner light surface with corrections 
up to $O(\varepsilon^2)$}

The solution with the 
corrections up to $O(\varepsilon^2)$ behaves near the inner light surface as  
\begin{eqnarray}
\psi= \psi^{(0)}_0+\varepsilon\psi^{(0)}_1+\varepsilon^2\psi^{(0)}_2+
\left(\psi^{(1)}_0+\varepsilon\psi^{(1)}_1\right)(y-1)+\psi^{(2)}_0(y-1)^2
+\cdots.
\end{eqnarray}
Although the solutions for $\psi^{(n)}_N$ for 
$(n,N)=(0,0),(0,1),(0,2),(1,0),(1,1),(2,0)$ have been derived in 
the previous sections,  
we again show them with a slightly different parameterization: 
\begin{eqnarray}
\psi^{(0)}_0&=&C^{(0)}_0(1-\cos\theta), \\
\psi^{(0)}_1&=&0, \\
\psi^{(0)}_2&=&\left[C^{(0)}_2\cos\theta\left(3\cos^2\theta-7\right)
+\frac{1}{4}C^{(0)}_0C'\left(\cos^2\theta-5\right)\right]\sin^2\theta, \\
\psi^{(1)}_0&=&C^{(0)}_0C'\sin^2\theta, \\
\psi^{(1)}_1&=&0, \\
\psi^{(2)}_0&=&\frac{1}{2}\left[3\left(C^{(0)}_0{C'}^2
-10C^{(0)}_2\right)\cos\theta+C^{(0)}_0C'\right]\sin^2\theta, 
\end{eqnarray}
where $C^{(2)}_0$ is given in this parameterization as 
\begin{equation}
C^{(2)}_0=\frac{3}{2}\left(C^{(0)}_0{C'}^2-10C^{(0)}_2\right). 
\end{equation}
By using the same parameterization as the above, the electric current is given by
\begin{eqnarray}
{\cal I}(X)&=&-\frac{1}{8\pi}\hat{X}\left(2-\hat{X}\right)\Bigl[2C^{(0)}_0
\nonumber \\
&-&\varepsilon^2\hat{X}\left(2-\hat{X}\right)
\left\{
C^{(0)}_0C'(1-\hat{X})+2C_2^{(0)}\left(2-18\hat{X}+9\hat{X}^2\right)
\right\}\Bigr],
\end{eqnarray} 
where $\hat{X}$ is defined by Eq.~(\ref{Xhat}). 
We see that the arbitrary constants are only $C^{(0)}_N$ ($N=0,2$) and $C'$ 
. The reason this result takes this form is because if we choose $\psi$ and $\partial_y\psi$  
on the event horizon such that the regularity conditions (\ref{HB}) and 
(\ref{HBD}) are satisfied, then $\psi$ and $\cal I$ are completely determined. 

\section{Summary and Discussion}

We studied a force-free magnetosphere in a static spherically symmetric black hole
spacetime with a degenerate event horizon.
We have found that if an electric current exists, higher multipole components of the magnetic 
field can be superposed upon the monopole component on 
the event horizon even if the two horizons degenerate into one horizon. 
This result is consistent with the numerical result given by Komissarov and McKinney: 
they showed that the magnetic field lines of higher multipole components 
can penetrate an extreme Kerr black hole 
if conductivity exists. The detailed geometrical structures of the 
extreme Kerr black hole and the extreme Reissner-Nordstr\"om 
black hole are different from each other. 
However, since the degenerate structures of the horizons of these black holes 
are similar, the present results may 
be applicable to a certain extent for the extreme Kerr black hole. 

If we require that there is no monopole component in the 
lowest-order configuration on the horizon, or equivalently, $\psi^{(0)}_0=0$, 
we obtain the trivial solution $\psi^{(0)}=0$, even though 
we take all-order corrections into account (see Appendix \ref{Proof}). 
Thus, the proposition in Appendix D seems to imply that 
there is no non-trivial configuration without a monopole component 
on the event horizon of the extreme Reissner-Nordstr\"om black hole, 
even if an electric current exists. 
But this is not necessarily true. 
In order to see this fact, note that 
there is an exact monopole solution for the Grad-Shafranov equation 
(\ref{eq:GS-eq}): 
\begin{equation}
\Psi=C(1-\cos\theta), 
\end{equation}
with the electric current
\begin{equation}
I=-\frac{\Omega_{\rm F}}{4\pi C} \Psi\left(2C-\Psi\right),
\end{equation}
where $C$ is an arbitrary constant. 
By contrast, the proposition in Appendix D 
implies that, if $\psi^{(0)}_0=0$, there is no higher-order correction by which the configuration of the perturbative solution 
on the event horizon approaches to the monopole 
configuration in our perturbation scheme.  
In other words, our perturbative solution with vanishing $\psi^{(0)}_0$ 
cannot approach to the above exact solution, 
even if we take into account all-order corrections. 
This fact suggests that even if an exact solution with 
a non-monopole configuration on the event horizon exists,  
the perturbative solution with vanishing $\psi^{(0)}_0$ 
cannot approach to such a solution in our perturbation scheme. 
This possibility 
may arise from the assumption for the electric current (\ref{assumptions}), 
which may be too strong, though the present analytic 
perturbation studies are impossible without this 
assumption. 

We would like to stress again that it is very difficult to obtain 
a stationary force-free magnetosphere by 
solving the Grad-Shafranov equation for the extreme black hole spacetime numerically. 
Thus, we need to invoke analytic methods, 
as in the present study, or numerical techniques to follow  
the dynamical evolution of a force-free Maxwell field until  
a stationary configuration is realized, as Komissarov and McKinney used. 
As discussed in this paper, 
in the case that there are two light surfaces 
in the extreme Reissner-Nordstr\"om black hole spacetime, 
even though the magnetic field is regular, 
both on the event horizon and inner light surface,  
it will be singular on the outer light surface.
If the angular velocity of the magnetic field is 
constant, two light surfaces necessarily exist.  
Thus, if the dynamical evolution of a force-free Maxwell field 
can be followed until it becomes stationary, then 
it is expected that the angular velocity 
decays far from the black hole 
so that the outer light surface does not exist.
The extremity of charge or angular momentum 
changes the structure of boundary conditions for the Grad-Shafranov equation and 
seems to strongly affect global structures of the black hole magnetosphere. 

Finally, we would like to suggest that 
analytic solutions obtained by this perturbation scheme becomes
a benchmark for a numerical scheme to obtain solutions for  
stationary configurations of astrophysical magnetospheres, since 
our perturbation scheme is also suitable for non-extreme black hole cases.  

\acknowledgments

The authors would like to thank M. Takahashi for useful lectures and discussions 
on black hole magnetospheres. 
C.Y. is supported by a JSPS Grant-in-Aid for Creative Scientific Research 
No. 19GS0219.

\appendix

\section{Derivation of Grad-Shafranov equation}
\label{Derivation}

A stationary, axisymmetric 
electromagnetic field implies $\partial_t A_a=0=\partial_\varphi A_a$. Then, 
from Eq.~(\ref{eq:F-def}) and the force-free condition (\ref{eq:ff-condition}), we have 
$F_{t\theta}/F_{\theta\varphi}=F_{tr}/F_{r\varphi}$. 
Using these equations, the components of $F_{ab}$ in the static coordinate system 
(\ref{metric}) are written in the form
\begin{equation}
F_{\mu\nu}=
\left(
\begin{array}{cccc}
0 & 0& \Omega_{\rm F}\partial_r A_\varphi &\Omega_{\rm F}\partial_\theta A_\varphi \\
0 & 0& -\partial_r A_\varphi & -\partial_\theta A_\varphi \\
-\Omega_{\rm F}\partial_r A_\varphi &\partial_r A_\varphi &0 & \sqrt{\gamma}B^\varphi\\
-\Omega_{\rm F}\partial_\theta A_\varphi & \partial_\theta A_\varphi & -\sqrt{\gamma}B^\varphi & 0
\end{array}
\right)~, \label{eq:F-comp}
\end{equation}
where $\Omega_{\rm F}$ is defined by Eq.~(\ref{eq:OmegaF-def}), and  
\begin{equation}
\gamma:=\frac{r^6\sin^2\theta}{\Delta}
\end{equation}
is the determinant of the intrinsic metric of the spacelike 
hypersurface labeled by $t$. 

Note that $\Omega_{\rm F}$ can be regarded as the angular velocity of the magnetic field. 
We consider an observer with an angular velocity 
$d\varphi/dt=\Omega_{\rm F}$. His or her 4-velocity is given by
$u^\mu = \Gamma (1, \Omega_{\rm F}, 0,0)$, where $\Gamma$ is a normalization factor. 
The electric field for this observer is given by $E_a=F_{ab}u^a$, and we can easily see 
from Eq.~(\ref{eq:F-comp}) that $E_a$ vanishes. Thus we may say that 
this observer is co-moving with the magnetic field, and the angular velocity 
of the magnetic field is $\Omega_{\rm F}$. 

Substituting Eq.~(\ref{eq:F-comp}) into the Jacobi identity $\partial_{[a}F_{bc]}=0$, we have
\begin{equation}
(\partial_r\Omega_{\rm F})\partial_\theta A_\varphi
-(\partial_\theta\Omega_{\rm F})\partial_r A_\varphi
=0. 
\end{equation}
The above equation implies that $\partial_a\Omega_{\rm F} \propto \partial_a A_\varphi$, 
or equivalently, the equi-$\Omega_{\rm F}$ surface agrees with the equi-$A_\varphi$ surface. 
Thus we have
\begin{equation}
\Omega_{\rm F}=\Omega_{\rm F}(A_\varphi). \label{eq:OmegaF}
\end{equation}

Using Eq.~(\ref{eq:F-comp}), the Maxwell equations imply the following equations: 
the $t$-component implies
\begin{equation}
\partial_r\left(r^2\Omega_{\rm F}\sin\theta
\partial_r A_\varphi\right)
+\partial_\theta\left(\frac{r^2\Omega_{\rm F}\sin\theta}{\Delta}
\partial_\theta A_\varphi\right)
=-4\pi\alpha\sqrt{\gamma}J^t; \label{eq:M-t-comp}
\end{equation}
the $\varphi$-component implies
\begin{eqnarray}
\partial_r\left(
\frac{\alpha^2}{\sin\theta}\partial_rA_\varphi
\right)
+\partial_\theta\left(
\frac{\alpha^2}{\Delta \sin\theta}
\partial_\theta A_\varphi
\right)
=-4\pi\alpha\sqrt{\gamma}J^\varphi; \label{eq:M-phi-comp}
\end{eqnarray}
the $r$-component implies
\begin{equation}
\partial_\theta(\alpha B_\varphi)=4\pi\alpha\sqrt{\gamma}J^r;
\label{eq:M-r-comp}
\end{equation}
and the $\theta$-component implies
\begin{equation}
\partial_r(\alpha B_\varphi)=-4\pi\alpha\sqrt{\gamma}J^\theta,
\label{eq:M-theta-comp}
\end{equation}
where
\begin{equation}
B_\varphi=r^2\sin^2\theta B^\varphi.
\end{equation}

Using Eq.~(\ref{eq:F-comp}), the force-free condition implies 
\begin{eqnarray}
J^r\partial_rA_\varphi+J^\theta\partial_\theta A_\varphi&=&0, 
\label{eq:F-t-comp}\\
(J^\varphi-J^t\Omega_{\rm F})\partial_rA_\varphi+\sqrt{\gamma}B^\varphi J^\theta&=&0, 
\label{eq:F-r-comp}\\
(J^\varphi-J^t\Omega_{\rm F})\partial_\theta A_\varphi-\sqrt{\gamma}B^\varphi J^r&=&0.
\label{eq:F-theta-comp}
\end{eqnarray}

Substituting Eq.~(\ref{eq:M-theta-comp}) to Eq.~(\ref{eq:F-r-comp}), and 
substituting Eq.~(\ref{eq:M-r-comp}) to Eq.~(\ref{eq:F-theta-comp}),  
we have
\begin{eqnarray}
(J^\varphi-J^t\Omega_{\rm F})\partial_rA_\varphi-\frac{1}{4\pi\alpha}
B^\varphi \partial_r(\alpha B_\varphi)&=&0, 
\label{eq:F-r-comp-dash}\\
(J^\varphi-J^t\Omega_{\rm F})\partial_\theta A_\varphi-\frac{1}{4\pi\alpha}
B^\varphi \partial_\theta(\alpha B_\varphi)&=&0. 
\label{eq:F-theta-comp-dash}
\end{eqnarray}
{}From the above equations, we have
\begin{equation}
\partial_r(\alpha B_\varphi)\partial_\theta A_\varphi
-\partial_\theta(\alpha B_\varphi)\partial_r A_\varphi
=0. 
\end{equation}
The above equations imply 
\begin{equation}
\alpha B_\varphi = {\cal B}(A_\varphi).
\end{equation}
Using the above equation, Eqs.~(\ref{eq:F-r-comp-dash}) and 
(\ref{eq:F-theta-comp-dash}) imply
\begin{equation}
J^\varphi-J^t\Omega_{\rm F}=\frac{1}{4\pi\alpha^2r^2\sin^2\theta}
{\cal B}\frac{d {\cal B}}{dA_\varphi}.
\label{eq:current}
\end{equation} 
{}From Eqs.~(\ref{eq:M-t-comp}) and (\ref{eq:M-phi-comp}), we have
\begin{eqnarray}
&&\partial_r \left(\frac{D\partial_r A_\varphi}{\sin\theta}\right)
+\frac{1}{\Delta}\partial_\theta\left(\frac{D\partial_\theta A_\varphi}
{\sin\theta}\right)
+\frac{r^2\Omega_{\rm F}\sin\theta}{\Delta}
\left[
\Delta(\partial_r A)\partial_r\Omega_{\rm F}
+(\partial_\theta A)\partial_\theta\Omega_{\rm F}
\right] 
\nonumber \\
&&=-4\pi\alpha\sqrt{\gamma}
\left(J^\varphi-J^t\Omega_{\rm F}\right),
\label{eq:M-tphi-comp}
\end{eqnarray}
where $D$ is defined by Eq.~(\ref{eq:D-def}). 
Noting that $\Omega_{\rm F}$ is a function of $A_\varphi$ and 
substituting Eq.~(\ref{eq:current}) into the right hand side 
of Eq.~(\ref{eq:M-tphi-comp}), we have
\begin{equation}
\partial_r^2A_\varphi+\frac{1}{\Delta}
\left(L_\theta A_\varphi+\frac{N}{D}\right)=0,
\label{eq:GS-equation}
\end{equation}
where
\begin{equation}
L_\theta A_\varphi:=\sin\theta\partial_\theta
\left(\frac{\partial_\theta A_\varphi}{\sin\theta}\right),
\end{equation}
and
\begin{equation}
N:=
\Delta(\partial_r A_\varphi)\partial_r D
+(\partial_\theta A_\varphi)\partial_\theta D
+\frac{r^2}{2}\sin^2\theta\frac{d\Omega_{\rm F}^2}{dA_\varphi}
\left[
\Delta(\partial_r A_\varphi)^2
+(\partial_\theta A_\varphi)^2\right]
+\frac{r^2}{2}\frac{d{\cal B}^2}{dA_\varphi}.
\end{equation}
The above equation is called the Grad-Shafranov equation. 

\section{Electric current and magnetic flux}
\label{I-Psi}

\begin{figure}
\centering
\includegraphics[width=0.5\textwidth]{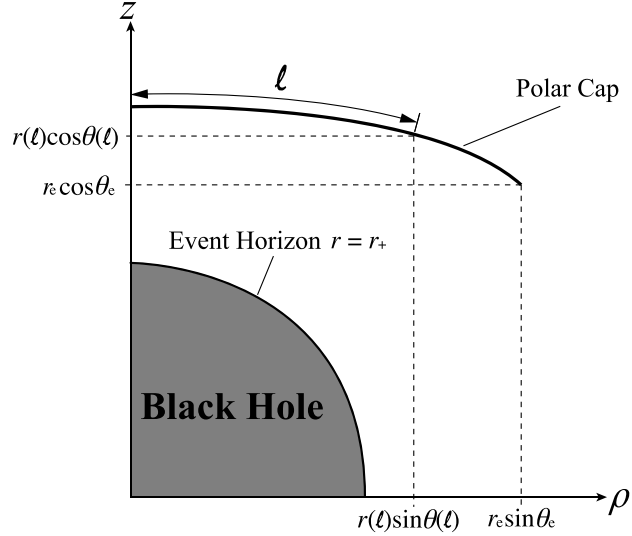}
\caption{
The schematic diagram of a polar cap which is an axisymmetric two-dimensional 
spacelike surface parameterized by the proper length $\ell$ measured from the symmetry 
axis $\theta=0$ along the polar cap. 
}
\label{fg:polar-cap}
\end{figure}

Here we introduce electric current $I$ and magnetic flux $\Psi$ on a 
spacelike hypersurface labeled by $t$ 
which penetrate downward and upward an axisymmetric polar cap, respectively. 
These quantities were first introduced by Macdonald and Thorne~\cite{Macdonald:1982}
and are related to ${\cal B}$ and $A_\varphi$ as follows.
The polar cap is parameterized by 
$\ell$ and $\varphi$, where $\ell$ is the proper length on the polar cap 
from $\theta=0$. The coordinates $r$ and $\theta$ on the polar cap are given as 
functions of $\ell$, i.e., $r=r(\ell)$ and $\theta=\theta(\ell)$: 
by definition, $\theta(0)=0$, and we 
assume that $r(0)>r_{+}$ (see Fig.~\ref{fg:polar-cap}). 
The orthonormal tangent vectors of the polar cap are 
\begin{eqnarray}
e_{(\ell)}{}^i&=&
\left(0,~\frac{dr}{d\ell},~\frac{d\theta}{d\ell}\right), \\
e_{(\varphi)}{}^i&=&
\left(\frac{1}{r\sin\theta},~0,~0\right).
\end{eqnarray}
Then, the upward unit normal to the polar cap is
\begin{equation}
n_i=\frac{r^2}{\sqrt{\Delta}}
\left(0,~\frac{d\theta}{d\ell},~-\frac{dr}{d\ell}\right).
\end{equation}

We assume that the edge of the polar cap is  
$r=r_{\rm e}$ and $\theta=\theta_{\rm e}$. 
Then, denoting the proper length 
$\ell$ at the edge by $\ell_{\rm e}$, we have
\begin{eqnarray}
I&=&-\int_0^{2\pi}\int_0^{\ell_{\rm e}}
\alpha J^ i n_i r\sin\theta d\ell d\varphi 
=-\frac{1}{2}\int_0^{\ell_{\rm e}}
\left[\frac{d\theta}{d\ell}\partial_\theta(\alpha B_\varphi)
+\frac{dr}{d\ell}\partial_r(\alpha B_\varphi)
\right]d\ell \nonumber \\
&=&-\frac{1}{2}\int_0^{\ell_{\rm e}}\frac{d (\alpha B_\varphi)}{d\ell} d\ell 
=-\frac{1}{2}\alpha B_\varphi\biggl|_{(r,\theta)=(r_{\rm e},\theta_{\rm e})},
\label{eq:I-def}
\end{eqnarray}
where we have used Eqs.~(\ref{eq:M-r-comp}) and (\ref{eq:M-theta-comp}) in 
the second equality and assumed that $B_\varphi|_{\theta=0}=0$ from the
regularity requirement. 
We can easily see from the above equation that the electric current 
through the polar cap is a function of the coordinate values of the edge, 
$(r_{\rm e},\theta_{\rm e})$. 
By a similar consideration to that for the electric current $I$, 
the magnetic flux $\Psi$ can be written in the form
\begin{eqnarray}
\Psi&=&
\int_0^{2\pi}\int_0^{\ell_{\rm e}}
\frac{1}{2}\epsilon^{ijk}F_{jk}n_ir\sin\theta d\ell d\varphi 
=\int_0^{2\pi}\int_0^{\ell_{\rm e}}
\frac{1}{\sqrt{\gamma}}
\left(F_{\theta\varphi}\frac{d\theta}{d\ell}
-F_{\varphi r}\frac{dr}{d\ell}\right)
\frac{r^3\sin\theta}{\sqrt{\Delta}} d\ell d\varphi \nonumber \\
&=&2\pi\int_0^{\ell_{\rm e}}
\left[\frac{d\theta}{d\ell}\partial_\theta A_\varphi
+\frac{dr}{d\ell}\partial_r A_\varphi\right] d\ell 
=2\pi\int_0^{\ell_{\rm e}}\frac{d A_\varphi}{d\ell} d\ell 
=2\pi A_\varphi |_{(r,\theta)=(r_{\rm e},\theta_{\rm e})}, 
\label{eq:Psi-def}
\end{eqnarray}
where $\epsilon^{ijk}$ ($\epsilon^{\varphi r\theta}=1/\sqrt{\gamma}$)
is the components of the 
skew tensor in the spacelike hypersurface labeled by $t$, and 
we have assumed $A_\varphi|_{\theta=0}=0$ from a regularity requirement. 
Thus we have
\begin{equation}
I=-\frac{1}{2}{\cal B}~~~~~~{\rm and}~~~~~~\Psi=2\pi A_\varphi.
\end{equation}
Rewriting Eq.~(\ref{eq:GS-equation}) using $I$ and $\Psi$, 
the Grad-Shafranov equation Eq.~(\ref{eq:GS-eq}) is obtained.

\section{Relation between the Kerr-Schild and 
Static coordinate systems}
\label{KS}

The line element of the Reisner-Nordstr\"om spacetime with the 
Kerr-Schild coordinate system $(T,\Phi,R,\Theta)$ is given by

\begin{eqnarray}
ds^2&=&-\frac{R^2}{R^2+2MR-Q^2}dT^2+R^2\sin^2\Theta d\Phi^2 \nonumber \\
&+&\frac{R^2+2MR-Q^2}{R^2}
\left(dR+\frac{2MR-Q^2}{R^2+2MR-Q^2}dT\right)^2+R^2 d\Theta^2,
\end{eqnarray}
where $M=r_++r_-$ and $Q^2=r_+r_-$. 
The relation between the Kerr-Schild coordinate system and 
the static one is given by
\begin{eqnarray}
dT&=&dt+\frac{2Mr-Q^2}{\Delta}dr, \\
d\Phi&=&d\varphi, \\
dR&=&dr, \\
d\Theta&=&d\theta.
\end{eqnarray}
{}From the above relation, we have $\Phi=\varphi$, $R=r$, and $\Theta=\theta$, and
\begin{eqnarray}
\frac{\partial}{\partial T}&=&\frac{\partial}{\partial t}, \\
\frac{\partial}{\partial \Phi}&=&\frac{\partial}{\partial \varphi}, \\
\frac{\partial}{\partial R}&=&-\frac{2Mr-Q^2}{\Delta}\frac{\partial}{\partial t}
+\frac{\partial}{\partial r}, \\
\frac{\partial}{\partial \Theta}&=&\frac{\partial}{\partial \theta}.
\end{eqnarray}
Using the above relations, we have 
\begin{equation}
A_\varphi=A_\Phi. 
\end{equation}
By virtue of the stationary, axisymmetric nature of the 
electromagnetic field, 
we can easily see that the components of $F_{ab}$ in the Kerr-Schild 
coordinate system are given as
\begin{eqnarray}
F_{T\Phi}&=&0,~~~~~~~~~~~~F_{TR}=\Omega_{\rm F}\partial_r A_\varphi,
~~~~F_{T\Theta}=\Omega_{\rm F}\partial_\theta A_\varphi, \\
F_{\Phi R}&=&-\partial_r A_\varphi,
~~~~F_{\Phi\Theta}=-\partial_\theta A_\varphi, \\
F_{R\Theta}&=&
\sqrt{\gamma}B^\varphi
-\frac{1}{\Delta}(2Mr-Q^2)\Omega_{\rm F}\partial_\theta A_\varphi \nonumber \\
&=&\frac{r^2}{\Delta\sin\theta}\left[{\cal B}
-\frac{1}{r^2}(2Mr-Q^2)\Omega_{\rm F}
\sin\theta\partial_\theta A_\varphi\right]. \label{eq:RT-comp} \\
\end{eqnarray}

It should be noted that all the ordinary derivatives of the Kerr-Schild 
coordinates are equivalent to those of the static coordinates 
for the stationary axisymmetric field $A_\Phi=A_\varphi$, 
\begin{eqnarray}
\partial_TA_\Phi&=&\partial_tA_\varphi=0,~~
\partial_\Phi A_\Phi=\partial_\varphi A_\varphi=0, \nonumber \\
\partial_RA_\Phi&=&\partial_rA_\varphi,~~
\partial_\Theta A_\Phi=\partial_\theta A_\varphi,~~
\partial_R^2A_\Phi=\partial_r^2A_\varphi,~~{\rm and}~~
\partial_\Theta^2 A_\Phi=\partial_\theta^2 A_\varphi.
\end{eqnarray}
Thus, even if the Kerr-Schild coordinate system is adopted, the 
equation for $A_\Phi$ takes exactly the same form as 
Eq.~(\ref{eq:GS-equation}). 

\section{Non-existence of non-monopole solution}
\label{Proof}

\noindent
{\bf Proposition}: Within the perturbation scheme developed in this paper, 
the solution for $\psi^{(0)}$ with vanishing lowest-order solution 
$\psi^{(0)}_0$ is the only trivial solution $\psi^{(0)}=0$. 

\vskip0.3cm
\noindent
{\it Proof.} We prove this by induction. We have already shown that, 
if we impose the non-existence of a monopole component for the 
lowest order of the perturbative solution, we obtain 
\begin{equation}
\psi^{(0)}_0=0. 
\end{equation}
Here, we assume that $\psi^{(0)}_N=0$ for $0\leq N\leq M$, or equivalently, 
\begin{equation}
\psi^{(0)}(\theta)=\varepsilon^{M+1}\sum_{N=0}^\infty \varepsilon^N
\psi_{N+M+1}(\theta).
\end{equation}
Then, we have
\begin{equation}
{\cal I}(\psi^{(0)})=\sum_{N=0}^\infty\varepsilon^N {\cal I}_N
\left(
\varepsilon^{M+1}\sum_{J=0}^\infty \varepsilon^J\psi_{J+M+1}
\right)
=\varepsilon^{M+1}{\cal I}_0{}'\psi^{(0)}_{M+1}+{\cal O}(\varepsilon^{M+2}),
\end{equation}
where
\begin{equation}
{\cal I}_0{}':=\frac{d{\cal I}_0(\psi)}{d\psi}\biggl|_{\psi=0},
\end{equation}
and we have used ${\cal I}_N(0)=0$, which is required from 
the regularity condition at $\theta=0$. 
Substituting the above equation into Eq.~(\ref{HB}), we 
obtain an equation of $O(\varepsilon^{M+1})$, 
\begin{equation}
{\cal I}_0{}'\psi^{(0)}_{M+1}
+\frac{1}{4\pi}\sin\theta\frac{d\psi^{(0)}_{M+1}}{d\theta}=0.
\end{equation}
Integrating the above equation, we have
\begin{equation}
\psi^{(0)}_{M+1}=C^{(0)}_{M+1}
\left[
\frac{\sin^2\theta}{(1+\cos\theta)^2}
\right]^{-2\pi {\cal I}_0{}'},
\end{equation}
where $C^{(0)}_{M+1}$ is an integration constant. 
The regularity condition implies $\psi^{(0)}_{M+1}|_{\theta=0}
=0=d\psi^{(0)}_{M+1}/d\theta|_{\theta=0}$. Hence, 
if $C^{(0)}_{M+1}$ does not vanish, 
\begin{equation}
-2\pi{\cal I}_0{}'=1+c^2
\end{equation}
must be satisfied, where $c$ is an arbitrary constant. 

In the neighborhood of $\theta=\pi$, 
$\sin\theta \sim \pi - \theta$ and $\cos\theta \sim -1+(\theta-\pi)^2/2$. 
Hence, if $C^{(0)}_{M+1}$ does not vanish, we have 
\begin{equation}
\lim_{\theta\rightarrow\pi}\psi^{(0)}_{M+1}=\infty.
\end{equation}
If we require the finiteness of $\psi^{(0)}_{M+1}$ at $\theta=\pi$, 
then $C^{(0)}_{M+1}$ must vanish, and, as a result, 
$\psi^{(0)}_{M+1}=0$ is obtained. \hfill Q.E.D.

\end{document}